\documentclass[12pt]{article}
\usepackage{amsmath, amsthm, amssymb}
\usepackage{graphicx}

\begin{document}

\title{Phase-space distribution functions for photon  propagation in waveguides coupled to a qubit}

\author{O. O. Chumak\footnote{Corresponding author: chumak@iop.kiev.ua} \, and E. V. Stolyarov \\[3mm]
 \\[5mm] Institute of Physics of the National Academy of Sciences\\ pr. Nauki 46, Kyiv-28, MSP 03028 Ukraine \\[3mm]\rule{0cm}{1pt}}

\maketitle 

\begin{abstract}

We investigate propagation of few-photon pulses in waveguides
coupled to a two-level system by means of the method of distribution
functions in coordinate-momentum space that provides a detailed
description of photon systems. We find that the distribution
function of the transmitted pulse can be negative for the
nonclassical input (i.e., single-photon Fock state). This reveals
the quasiprobability nature of photon distribution. Analytical
expressions for photon densities in the momentum space as well as in
the coordinate space are obtained for the mentioned single-photon
Gaussian input. We also study evolution of the multimode
coherent-state input for an arbitrary photon number. Time-dependent
differential equations describing average densities and fluctuations
of outgoing photons are derived and solved. Influence of the number
of input photons, pulse width, and radiation-atom interaction
strength on the statistical properties of the fluctuations is
investigated.

\end{abstract}

\section{INTRODUCTION}

Traditionally the model of two-level system (TLS) is used to
describe the interaction of electromagnetic radiation with atoms
\cite{mand}. This model is quite reasonable if radiation frequency
and transition frequency of the corresponding two levels are very
close to each other. It was suggested to utilize Rydberg atoms
controlled by electromagnetic fields as  qubits in quantum
information technologies \cite{s3}.  There are other implementations
of the qubits. Trapped ions \cite{s1},\cite{s2}, semiconductor
quantum dots \cite{s4}, superconducting Josephson junctions can be
used for this purpose. Qubits based on the Josephson junctions are
recognized now to be the most promising for realization of quantum
information processing devices (see, for example, Refs.
\cite{urb}-\cite{wallr}). There is a technological opportunity to
couple  qubits via transmission lines. Individual photons can act as
transmitters of quantum states between remote qubits.

The aforesaid illustrates motivations to study TLS coupled to
transmission lines. Atoms having a large dipole moment (Rydberg
atoms) as well as transmission lines (including optical waveguides)
that concentrate radiation energy in small volumes are used to
increase the coupling. Strong interaction is desired for many
applications whose aim is to achieve effective influence of one
subsystem on the other. At the same time increase of interaction
results in  more pronounced nonlinearity of the system. This
complicates theoretical analysis. Therefore many theoretical results
were obtained only numerically (see, for example,
\cite{dro}-\cite{buz}). Fortunately, analysis can be simplified
considerably for some particular states of the system. First of all,
a single-photon Fock state of the  incident radiation should be
mentioned (see, for example, recent papers \cite{she}-\cite{fan}).
Also things get simplified if the incident light is in a coherent
state. For example, Ref. \cite{fan} deals with  radiation which is
initially in a single-mode coherent state. Much earlier paper
\cite{dom} considers more general and more important for
applications multimode coherent-state pulses. Results of Ref.
\cite{dom} gives a possibility not only to obtain the reflectance
and transmittance of a wave packet but also to study spatial
structure of outgoing radiation and its dependence on the incident
pulse shape. Moreover, paper \cite{dom} describes effective
photon-photon "interaction" induced by coupling of the radiation
with atoms.

Different formalisms are used in the cited papers. Method of
scattering matrix \cite{shi} which is equivalent to the input-output
formalism \cite{wal} is applied to photon scattering by TLS in
\cite{fan}. Authors of \cite{dom} use an alternative approach based
on calculation of Poynting vectors to study  similar physical
systems.

Recently we have applied method of photon phase-space distribution
function to the problem of light propagation in the Earth atmosphere
\cite{ber}, \cite{ber1}. In the present paper we use this method for
description of light propagation in waveguides. We obtain a spatial
structure and spectrum of the transmitted and reflected radiation
which are useful for design of radiation with desired properties. Besides we analyze
physical nature of the phase-space distribution functions. It is
shown that they, like $Q-$ or $W-$distributions, can be negative for
some specific parameters of the incident radiation.

Also equations describing fluctuations of outgoing photons are
derived and solved. We show that the variance of the reflected
radiation may be essentially lower than that of a coherent-state
pulse. Thus, few-photon pulses with favorable statistical properties
can be generated in course of radiation-TLS interaction.

In the next Section, one-dimensional distribution functions are
defined in terms of creation and annihilation operators of waveguide
modes. The standard Hamiltonian describing light propagation and
interaction with TLS is used to derive evolution equations.

\section{HAMILTONIAN AND PHOTON DISTRIBUTION FUNCTIONS}

We consider a model Hamiltonian describing a two-level atom coupled
to a single-polarization waveguide. The waveguide modes are assumed
to form a one-dimensional continuum. Then the Hamiltonian is given
by ($\hbar=1$)
\begin{equation}\label{one}
 H = \int dk(\omega_k^ll_k^\dag l_k + \omega_k^rr_k^\dag r_k)+
 \frac {\omega_a}2\sigma_z+g\int dk\big [\sigma_+(l_k+ r_k)+
 (l_k^\dag +r_k^\dag)\sigma_-\big],
 \end{equation}
where $l_k$ and $r_k$ are the annihilation operators of photons
propagating from the left side to the right side and vice versa,
respectively. Photon frequencies are denoted correspondingly by
$\omega_k^{l,r}$. Notations $l_k^\dag$ and $r_k^\dag$ stand for the
creation operators. For a symmetric waveguide the linearized in $k$
dispersions in the vicinity of $\omega^{l,r}=\omega_0$ are given by
$\omega_k^{l,r }=\omega_0\pm vk$ where  $v$ and $-v$ ($v>0$) are
velocities of waves propagating from the left and from the right,
respectively. Atomic operators $\sigma _z$ and $\sigma _\pm$ are
defined by Pauli matrices: $\sigma_{\pm}=\frac 12(\sigma_x\pm
i\sigma_y)$, $\sigma_+\sigma_-=(\sigma_z+1)/2$.

Field variables follow the usual bosonic commutation rules
\[ [l_k,l_{k^\prime} ^\dag]=[r_k,r_{k^\prime}^\dag]=\delta (k-k^\prime),\]
while the rest of commutators vanish. Also,  field variables commute
with atomic variables.

The  first term in the right side of Eq. \ref{one} describes
electromagnetic field in the waveguide. The second term is the
Hamiltonian of a two-level atom with transition frequency
$\omega_a$. The third term describes the radiation-atom interaction
whose strength is determined by parameter $g$. It is assumed that
the atom is positioned  at the origin of the coordinate system which
makes the Hamiltonian to be independent explicitly of the atom
coordinate. The interaction is presented in the rotating-wave
approximation. Recently an approach which is free of this widely
used constraint is developed in Ref. \cite{bez}.

Photons moving from the left can be described by their density in
the phase space ($x,q$-space). The corresponding function is defined
as
\begin{equation}\label{two}
 f^l(x,q,t)=\frac 1{2\pi}\int dke^{-ikx}l^\dag _{q+k/2}l_{q-k/2},
 \end{equation}
where all operators are given in the Heisenberg picture. The
distribution function (\ref{two}) is defined  by analogy with the 3D
case (see more details in Ref. \cite{sus}). By integrating
(\ref{two}) over $x$ we obtain the density of $l$-photons in the
momentum space:
\begin{equation}\label{thr}
\hat{n}^l(q,t)\equiv \int dxf^l(x,q,t)=l^\dag _{q}(t)l_{q}(t).
\end{equation}
 Accounting for the linear
dependence $\omega^l(k)$ we can conclude that the average value
$\langle \hat{n}^l(q,t)\rangle$ determines the spectral distribution
of photons moving from  the left. The spectrum can be obtained from
Eq. (\ref{thr}) by changing $q\rightarrow(\omega-\omega_0)/v$.

Similarly we can express the photon density  in the coordinate
space, $\hat{\rho}_l(x,t)$, in terms of the distribution function,
$f^l(x,q,t)$, as:
\begin{equation}\label{fou}
\hat{\rho}_l(x,t)\equiv\int dqf^l(x,q,t)=\frac 1{2\pi}\int
dqdke^{-ikx} l^\dag _{q+k/2}l_{q-k/2}.
\end{equation}
Furthermore, by integrating  $\hat{\rho}_l(x,t)$ over $x$ in the
range of localization of the transmitted pulse we obtain the
operator of total number of the transmitted photons, $N_l$, as
\begin{equation}\label{fiv}
\hat{N}_l(t)=\int dx\hat{\rho}_l(x,t)=\int dql^\dag _q(t)l_q(t).
\end{equation}
Expression (\ref{fiv}) can be used for obtaining both transmittance
and fluctuations of the transmitted photons.

Similar relationships for the $r$-photons follow from Eqs.
(\ref{two})-(\ref{fiv}) by replacing $l\rightarrow r$.
Characteristics of the outgoing radiation depend on the initial
state of the system and on the evolution of the above-mentioned
operators.

\section{EVOLUTION EQUATIONS}

Evolution of the system variables is governed by a set of coupled
Heisenberg  equations
\begin{equation}\label{six}
(\partial_t+i\omega_q^l)l_q=-ig\sigma_-,
\end{equation}
\begin{equation}\label{sev}
(\partial_t+i\omega_q^r)r_q=-ig\sigma_-,
\end{equation}
\begin{equation}\label{eig}
(\partial_t+i\omega_a)\sigma_-=ig\sigma_z\int dq(l_q+r_q),
\end{equation}
\begin{equation}\label{nin}
\partial_t\sigma_z=-i2g\int
dq[\sigma_+(l_q+r_q)-(l_q^++r_q^+)\sigma_-].
\end{equation}
Equations for variables $l_q^\dag ,r_q^\dag , \sigma_+$ can be
obtained by Hermitian conjugation of Eqs. (\ref{six})-(\ref{eig}).

Following Ref. \cite{wal} we represent a formal solution of Eq.
(\ref{six}) as
\begin{equation}\label{ten}
l_q(t)=\tilde{l}_q(t)-ig\int_{t_0}^tdt^\prime
e^{-i\omega_q^l(t-t^\prime)}\sigma_-(t^\prime),
\end{equation}
where $\tilde{l}_q(t)=l_q(t_0)e^{-i\omega_q^l(t-t_0)}$ and $t>t_0$.
It is assumed that the pulse,  localized at $t=t_0$ on the left from
the atom, moves to the right. The first term in the right side of
Eq. (\ref{ten}) describes the free-field propagation, while the
second one represents the  atom radiation. By integrating Eq.
(\ref{ten}) over $q$, we obtain a useful relationship
\begin{equation}\label{ele}
\int dql_q(t)=\int dq\tilde{l}_q(t)-\frac {i\pi g}v\sigma_-(t),
\end{equation}
which is widely used in the literature. Then evolution of the atomic
operators is governed by the equations
\begin{equation}\label{tve}
(\partial_t+i\omega_a+\Gamma/2)\sigma_-=ig\sigma_z\int
dq(\tilde{l}_q+\tilde{r}_q),
\end{equation}
\begin{equation}\label{thir}
(\partial_t+\Gamma)(\sigma_z+1)=-i2g\int
dq[\sigma_+(\tilde{l}_q+\tilde{r}_q)-(\tilde{l}_q^++\tilde{r}_q^+)\sigma_-],
\end{equation}
where $\Gamma=4\pi g^2/v$ and the tilde means the dependence on $t$
similar to the dependence $\tilde{l}_q(t)$. If we consider the
tilded variables as given functions then we have a closed set of
linear equations for obtaining atomic variables
$\sigma_\pm(t),\sigma_z(t)$. Using Eq. (\ref{tve}) we can exclude
variables $\sigma_\pm$ from Eq. (\ref{thir}). When
$\Gamma(t-t_0)\gg1$ we get
\begin{equation}\label{four}
(\partial_t+\Gamma)(\sigma_z+1)
\end{equation}
\[=-2g^2\int dq
dk\int_{t_0}^tdt^\prime[e^{(i\omega_a-\Gamma/2)(t-t^\prime)}(\tilde{l}^\dag_k+
\tilde{r}^\dag_k)_{t^\prime}\sigma_z(t^\prime)(\tilde{l}_q+
\tilde{r}_q)_t+H.c.].\]
 The distribution function and the atom
variables $\sigma_\pm(t)$ are related by \[f^l(x,q,t)=\frac
1{2\pi}\int dke^{-ikx}[\tilde{l}^\dag_{q+k/2}(t)+
 ig\int_{t_0}^tdt^\prime
e^{i\omega_{q+k/2}^l(t-t^\prime)}\sigma_+(t^\prime)]\]
\begin{equation}\label{fift}
\times[\tilde{l}
_{q-k/2}(t)-
 ig\int_{t_0}^tdt^\prime
e^{-i\omega_{q-k/2}^l(t-t^\prime)}\sigma_-(t^\prime)].
 \end{equation}
Eq. (\ref{fift}) follows directly from Eqs. (\ref{two}) and
(\ref{ten}). By integrating over $q$, we obtain the expression for
the photon density
\begin{equation}\label{fift1}
\hat{\rho}_l(x,t)=\tilde{\rho}_l(x,t)+\frac
\Gamma{4v}\Sigma_{t-x/v}+i\frac gv\int
dq(\sigma_+\tilde{l}_q-\tilde{l}_q^\dagger\sigma_-)_{t-x/v},
 \end{equation}
where $\Sigma\equiv\sigma_z+1$, $x>0$ and $\tilde{\rho}_l(x,t)$ is
presented in terms of the "free" operators $\tilde{l}(t)^\dagger$
and $\tilde{l}(t)$. The reflected photons can be described by the
operator
\begin{equation}\label{fift2}
\hat{\rho}_r(x,t)=\tilde{\rho}_r(x,t)+\frac
\Gamma{4v}\Sigma_{t+x/v}+i\frac gv\int
dq(\sigma_+\tilde{r}_q-\tilde{r}_q^\dagger\sigma_-)_{t+x/v},
 \end{equation}
where $x<0$ and $\tilde{\rho}_r(x,t)$ is defined via
$\tilde{r}(t)^\dagger $ and $\tilde{r}(t)$. In what follows we will
omit the term $\tilde{\rho}_r(x,t)$ because of the absence of
photons propagating from the right at $t=t_0$.

\section{SINGLE-PHOTON FOCK STATE}

We consider the simplest situation when the incident Gaussian wave
packet contains only one photon distributed among waveguide modes.
When this photon propagates from the left the single-photon Fock
state can be defined as \cite{mand}
\begin{equation}\label{sixt}
|1_l\rangle=\frac {w^{1/2}}{\pi^{1/4}}\int
dke^{-ikx_0}e^{-k^2w^2/2}l^\dag_k(t_0)|0\rangle ,
 \end{equation}
 where $|0\rangle$ is the vacuum state of the system. The coefficient before the
 integral is the normalization constant.

The average value of the initial distribution function  is given by
\begin{equation}\label{seve}
\langle 1_l|f^l(x,q,t_0)|1_l\rangle=\frac 1\pi
e^{-(x-x_0)^2/w^2}e^{-q^2w^2} .
 \end{equation}
It follows from Eq. (\ref{seve})  that $w$ can be interpreted as the
width of a pulse centered  at $x=x_0$. Moreover, the pulse spread in
the momentum space, $\Delta q$, is of the order of $1/w$. Before
photons reach the ground-state atom their distribution function
evolves as
\begin{equation}\label{eite}
\langle 1_l|f^l(x,q,t)|1_l\rangle=\frac 1\pi
e^{-X^2(t)/w^2}e^{-q^2w^2} ,
 \end{equation}
where $X(t)=x-x_0-v(t-t_0)$. To obtain the average $\langle
f^l(x,q,t)\rangle$ in the domain $v(t-t_0)>-x_0$ we should use its
general form ($\ref{fift}$). Simple calculations result in the
following average distribution function of the transmitted signal:
\begin{equation}\label{nine}
\langle 1_l|f^l(x,q,t)|1_l\rangle=\frac w{2\pi^{3/2}}\int dk
e^{-ikX(t)}e^{-(q^2+k^2/4)w^2}
\end{equation}
\[\times\bigg[1-\frac \Gamma 2\frac
{\Gamma/2+ikv}{(\omega_{a0}-qv)^2-(kv-i\Gamma)^2/4} \bigg ],\] where
$\omega_{a0}=\omega_a-\omega_0$.

For obtaining Eq. (\ref{nine}) the relations
\begin{equation}\label{tvan}
\sigma_-(t)|1_l\rangle =\big[\langle 1_l|\sigma_+(t)\big]^\dag
\end{equation} \[
=-ig\bigg(\frac {4\pi}{w^2}\bigg)^{1/4}
e^{-i\omega_0(t-t_0)}\int_{t_0}^tdt^\prime
e^{-(i\omega_{a0}+\Gamma/2)(t-t^\prime)}
e^{-[x_0+v(t^\prime-t_0)]^2/2w^2}|0\rangle\] are used. Also it is
assumed that the atom has a sufficient time to relax to the ground
state. It is so if
\begin{equation}\label{a}
v(t-t_0)+x_0\gg w+v/\Gamma.
\end{equation}

By integrating Eq. (\ref{nine}) over $x$ we get
\begin{equation}\label{tvon}
\langle \hat{n}^l(q,t)\rangle =\frac w{\pi^{1/2}}
e^{-q^2w^2}\bigg[1- \frac
{\Gamma^2/4}{(\omega_{a0}-qv)^2+\Gamma^2/4} \bigg ].
\end{equation}
\begin{figure}[!ht]
\centering
\includegraphics{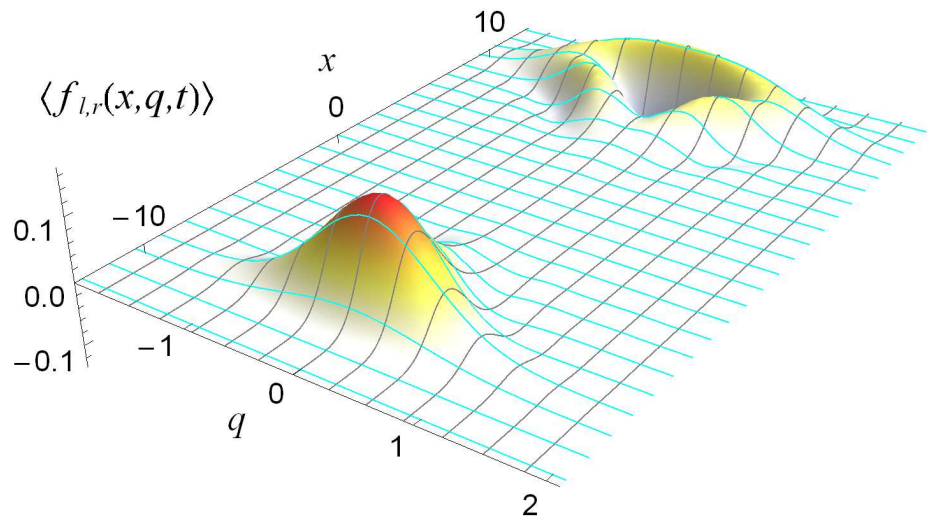}
\caption{(Color online) Photon  distributions  $\langle
f^{l}(x,q,t)\rangle$ ($x>0$) and $\langle f^{r}(x,q,t)\rangle$
($x<0$) as  functions of the phase-space variables, $x$ and $q$.
Calculations are performed for $t=20,\,t_0=0,\,\Gamma=1$,$\,
\omega_{a0}=0$, and $x_0=-10$. Quantities $x,q,t$, and $\Gamma$ are
given in units of $w$ (pulse width), $w^{-1}$ (inverse pulse
width), $w/v$ (pulse duration), and $v/w$ (inverse pulse
duration), respectively.}
\end{figure}
 The second term in the square brackets describes
resonant reflection of the waves. This process is efficient when
$\omega_{a0}-vq\equiv\omega_a-\omega^l_q\leq\Gamma/2.$  If
$\omega_{a0}\neq 0$ the spectrum of the transmitted field is
asymmetric which is different from the spectrum of the incident
radiation. In fact, Eq. (\ref{tvon}) describes  filtering properties
of the atom. It is  straightforward to generalize Eq. (\ref{tvon})
for an arbitrary shape of the incident pulse.

The average photon density is
\begin{equation}\label{tvtw}
\langle \hat{\rho}_l(x,t)\rangle =\frac 1{\pi^{1/2}w} \bigg
|e^{-X^2(t)/2w^2}-\frac {i\Gamma w}{2^{3/2}\pi^{1/2}}\Phi
[X(t)]\bigg|^2 ,
\end{equation}
where \[\Phi(x)=\int dq\frac{e^{-iqx-q^2w^2/2}}{\omega_{a0}-qv+i
\Gamma/2}.\] For a very short incident pulse, $(\Gamma w/2v)<<1$,
contribution of the term with $\Phi$ is negligible regardless of the
value of $\omega_{a0}$. This means that reflection is small for this
case.

In the opposite case, $(\Gamma w/2v)\gg1$, we obtain the photon
density
\begin{equation}\label{tvth}
\langle \hat{\rho}_l(x,t)\rangle =\frac 1{\pi^{1/2}w}
e^{-X^2(t)/w^2}\bigg|1-(1-i2\omega_{a0}/\Gamma)^{-1}\bigg|^2,
\end{equation}
which is equal to zero when $\omega_{a0}=0$. Hence, this is the case
of full reflection. Nevertheless, for large detuning,
$2\omega_{a0}/\Gamma\gg1$, the radiation-atom interaction vanishes
 resulting in almost full transmission.

The distribution function of the reflected radiation is given by
\begin{equation}\label{tvfo}
\langle 1_l|f^r(x,q,t)|1_l\rangle=\frac {\Gamma^2w}{8\pi^{3/2}}\int
dk e^{-ikX^r(t)}\frac {e^{-(q^2+k^2/4)w^2}}
{(\omega_{a0}+qv)^2-(kv+i\Gamma)^2/4},
\end{equation}
 where
$X^r(t)=x+x_0+v(t-t_0)$.

Typical distribution functions, $\langle f^{l,r}(x,q,t)\rangle$, are
shown in Fig. 1. In contrast to the initial positive distribution
(\ref{seve}) the region with negative values of $\langle
f^{r}(x,q,t)\rangle$ can be seen here that indicates a nonclassical
nature of the reflected radiation. Physical quantity  $\langle
f^{l,r}(x,q,t)\rangle$ can be interpreted as a quasiprobability
rather than a probability of the photon distribution in the phase
space.
\begin{figure}[!ht]
\centering
\includegraphics{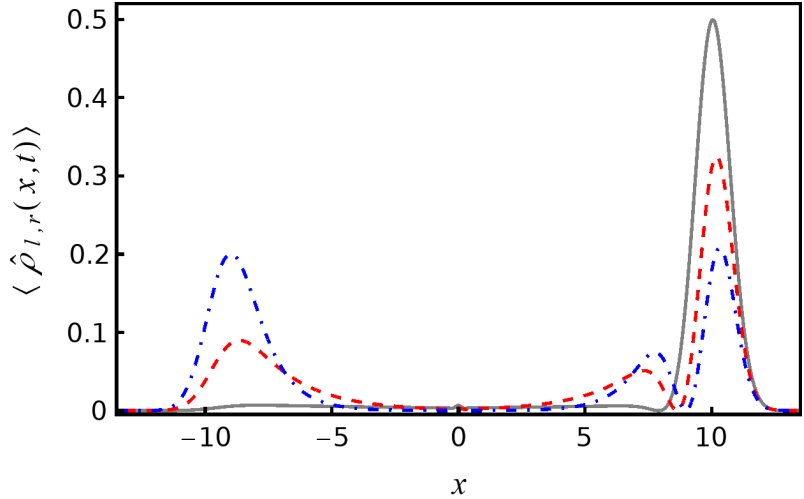}
\caption{(Color online) Photon density in the configuration-space.
Gray, red dashed, and blue dot-dashed curves are shown for $\Gamma$
equal to $0.1, 0.5$ , and $1$, respectively. Other parameters are as
in Fig. 1.}
\end{figure}

  Using Eq. \ref{tvfo} we obtain coordinate and momentum
distributions of the reflected photons as
\begin{equation}\label{tvfi}
\langle \hat{\rho}_r(x,t)\rangle =\frac {\Gamma^2w}{8\pi^{3/2}}
\bigg|\Phi[-X^r(t)]\bigg|^2 ,\quad\langle \hat{n}^r(q,t)\rangle
=\frac {\Gamma^2w}{4\pi^{1/2}} \frac {e^{-q^2w^2}}
{(\omega_{a0}+qv)^2+\Gamma^2/4} .
\end{equation}
The second expression in (\ref{tvfi}) shows the resonant character
of the reflection at $\omega_a-\omega_q^r\sim \Gamma/2$.

As we see from Eqs. (\ref{tvon}), (\ref{tvtw}) and (\ref{tvfi})
there are no regions with negative values of photon distributions
$\langle \hat{\rho}_{r,l}(x,t)\rangle$ and $\langle
\hat{n}^{r,l}(q,t)\rangle$. A set of curves in Figs. 2, 3 shows the
expected tendency: reflection is bigger for stronger interaction.
\begin{figure}[!ht] \centering
\includegraphics{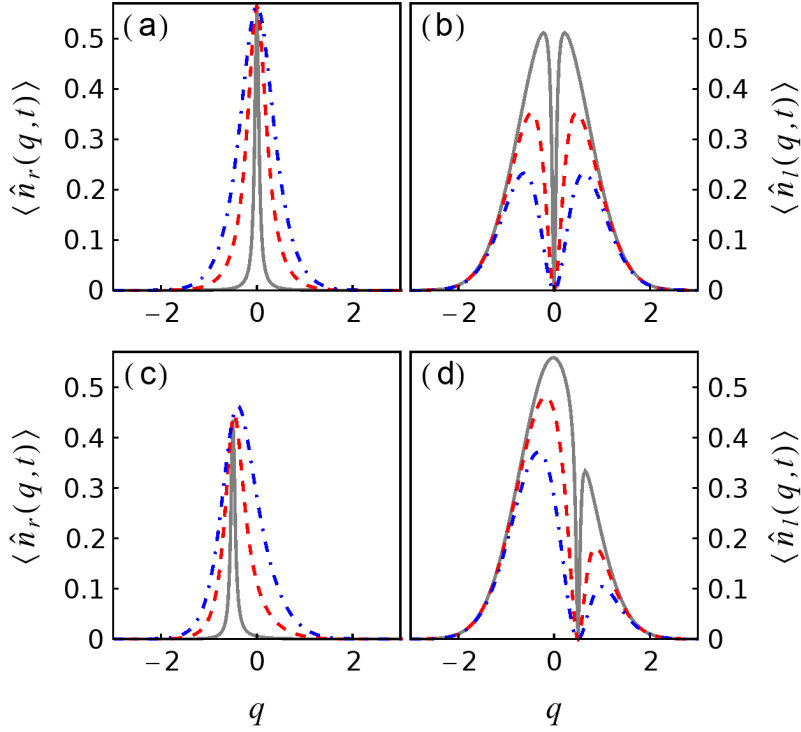}
\caption{(Color online) Photon distribution in the momentum space.
$\omega_{a0}=0$ (a,b); $\omega_{a0}=0.5$ (c,d). $\omega_{a0}$ is
given in units of $v/w $. Other notations are as in Figs. 1,2.}
\end{figure}

 Asymmetry of curves with respect to the central point, $q=0$, is
 seen in Figs. 3c and 3d.
This is because only the incident photons with $q>0$ can be in
resonance with the atom. Hence, they have the biggest probability to
be reflected thus forming pronounced minima in $\langle
\hat{n}^{l}(q,t)\rangle$ curves and the corresponding maxima in
$\langle \hat{n}^{r}(q,t)\rangle$ curves. In the case of negative
values of $\omega_{ao}$ similar plots can be obtained by formal
replacement $q\rightarrow-q$ in Figs. 3c and 3d.

\section{COHERENT-STATE OF THE INCIDENT RADIATION}

Incident Gaussian pulse can be represented by a coherent-state wave
packet. Following the paper \cite{blo} we define the corresponding
wave function as
\begin{equation}\label{tvsi}
 |\Psi\{\alpha\}\rangle=\exp\bigg \{\int dk[\alpha_kl^\dag_k-
 \alpha^*_kl_k]\bigg
 \}|0\rangle,
\end{equation}
where \[\alpha_k=\pi^{-1/4}(N_0w)^{1/2}e^{-ikx_0-k^2w^2/2}.\] It can
be easily verified that  function (\ref{tvsi}) is the eigenfunction
of all annihilation operators:
$l_k|\Psi\{\alpha\}\rangle=\alpha_k|\Psi\{\alpha\}\rangle$. We use
this property in further analysis.

By averaging the initial distribution function over the state
(\ref{tvsi}) we obtain
\begin{equation}\label{tvse}
\langle\Psi\{\alpha\}|f^l(x,q,t_0)|\Psi\{\alpha\}\rangle =\frac
{N_0}\pi e^{-(x-x_0)^2/w^2}e^{-q^2w^2},
\end{equation}
 which is very similar to Eq. (\ref{seve}). The only free parameter, $N_0$,
equal to the average number of photons per pulse, differs
 Eq. (\ref{tvse}) from Eq. (\ref{seve}).

We  use Eq. (\ref{fift}) to study photon density of the reflected
and transmitted radiation. Integrating Eq. (\ref{fift}) over $q$ and
using Eq. (\ref{four}),  the average density of photons,
$\hat{\rho}_l(x,t)$  is obtained as
\begin{equation}\label{tvei}
\langle \hat{\rho}_l(x,t)\rangle =\langle \tilde{\rho}_l(x,t)\rangle
-\frac{\Gamma}{4v}\langle \Sigma\rangle_{t-x/v}-\frac
1{2v}\partial_t \langle \Sigma\rangle_{t-x/v}.
\end{equation}
The last two terms in Eq. (\ref{tvei}) describe atom response and
 interference of the response with the incoming field, respectively.
 A similar term for the backward-propagating pulse is given by
\begin{equation}\label{tvni}
\langle \hat{\rho}_r(x<0,t)\rangle = \frac{\Gamma}{4v}\langle
\Sigma\rangle_{t+x/v}.
\end{equation}
Eq. (\ref{tvni}) describes the radiation back-scattered by the atom.
 As we see the field distribution in the waveguide is expressed in
 terms of the average
 $\langle \Sigma\rangle$ which describes an atomic state.  After
 averaging (\ref{four}) over the
 initial wave function, $|\Psi\{\alpha\}\rangle$, we get
\begin{equation}\label{thh}
(\partial_t+\Gamma )\langle \Sigma\rangle_t
=-4g^2p(t)\int_{t_0}^tdt^\prime e^{-\Gamma
(t-t^\prime)/2}p(t^\prime)\langle \sigma_z\rangle_{t^\prime}
\cos[\omega_{a0}(t-t^\prime )],
\end{equation}
where $p(t)=\pi^{1/4}\bigg(\frac{2N_0}w \bigg
)^{1/2}\exp\{-[x_0+v(t-t_0)]^2/{}2w^2\}$. Eq. (\ref{thh}) should be
completed with the initial condition $\langle
\Sigma\rangle_{t_0}=0$.

The integro-differential equation (\ref{thh}) can be transformed
into a differential equation. We consider the simplest case of
$\omega_{a0}=0$. Applying operator $\partial_t$ to both parts of Eq.
(\ref{thh}) we obtain
\begin{equation}\label{toh}
\hat{L}_t\langle \Sigma\rangle =4g^2p^2(t),
\end{equation}
where \[\hat{L}_t=\partial^2_t +\bigg[\frac
32\Gamma+(t-t_e)\frac{v^2}{w^2}\bigg]\partial_t+\bigg[
\frac{\Gamma^2}{2}+(t-t_e)\frac{v^2}{w^2}\Gamma+4g^2p^2(t)\bigg]\]
and $t_e=t_0+|x_0|/v$.
\begin{figure}[!ht]
\centering
\includegraphics{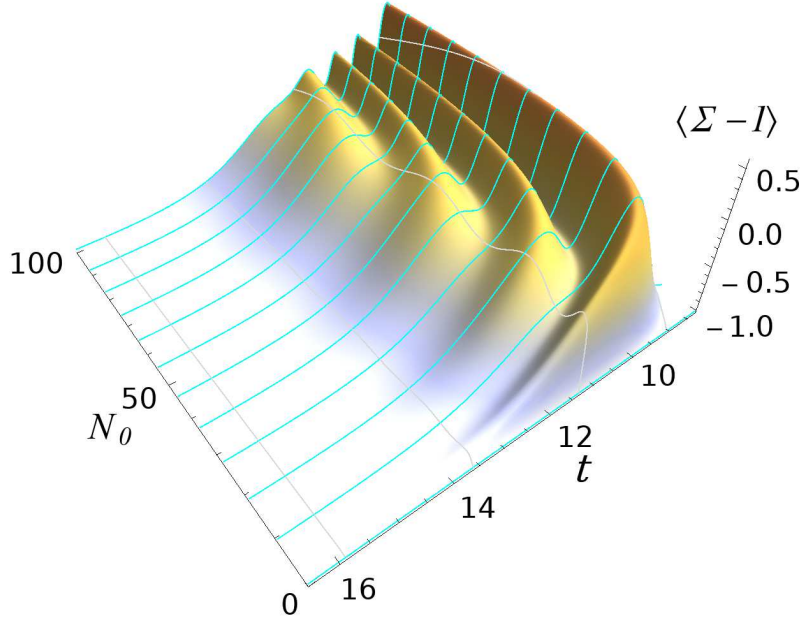}
\caption{(Color online) Atom excitation dynamics vs initial photon
number $N_0$ for $\Gamma=1$.}
\end{figure}

In the case of a long pulse, $(\Gamma w/2v)\gg 1$, the
quasistationary state of $\langle\Sigma\rangle$ given by
\[\langle \Sigma\rangle_{qs}\approx\bigg(
1+\Gamma^2/8g^2p^2(t)\bigg)^{-1},\] can be realized. It follows from
the above expression that for  large (small) driving fields,
$p^2(t)\rightarrow\infty \,( p^2(t)\rightarrow 0$), the value of
$\langle \Sigma\rangle_{qs}$ is equal to $1\, (0 )$ in agreement
with the simplest qualitative reasonings. There are damped
oscillations of $\Sigma $ around this state. Their evolution is
governed by Eq. (\ref{toh}) which in the long-pulse limit reduces to
\begin{equation}\label{tth}
\delta \ddot{\Sigma}+\delta\dot{\Sigma}\frac 32\Gamma+ \delta \Sigma
\bigg[\frac {\Gamma^2}2+4g^2p^2(t)\bigg]=0,
\end{equation}
where $\delta\Sigma=\langle \Sigma\rangle -\langle
\Sigma\rangle_{qs}$. Ignoring the dependence of $p$ on $t$ we seek a
solution in the form $\delta\Sigma\sim e^{\lambda t}$. Then the
equation for  $\lambda$ is given by
\begin{equation}\label{tthh}
\lambda^2+\lambda\frac32\Gamma+\frac {\Gamma^2}2+4g^2p^2(t)=0.
\end{equation}
It follows from Eq. (\ref{tthh}) that oscillations of $\langle
\Sigma(t)\rangle$ arise if only $(2vp/\pi g)>1$,oscillation decay
rate with the oscillation decay rate of the order of $3\Gamma/4$.

\begin{figure}[!ht]
\centering
\includegraphics[width=0.7\textwidth]{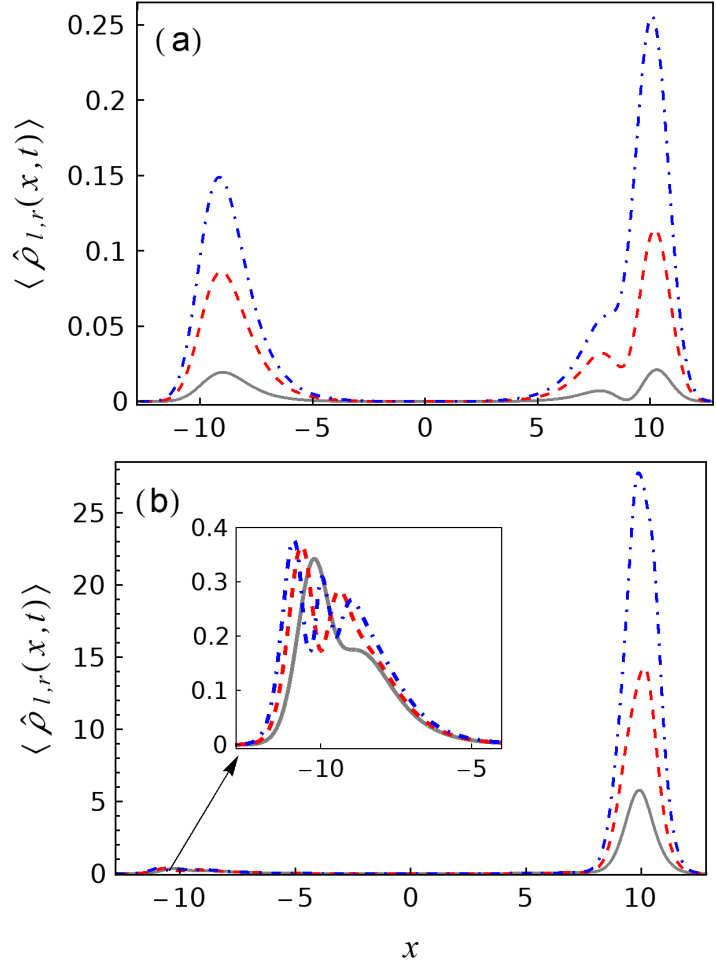}
\caption{(Color online) Photon configuration-space densities:   (a)
$N_0 = 0.1$ - gray solid line, $N_0 = 0.5$ - red dashed line, $N_0 =
1$ - blue dash-dot line; (b) $N_0 = 10$ - gray solid line, $N_0 =
25$ - red dashed line, $N_0 = 50$ - blue dash-dot line (reflected
pulses are shown in the inset). Input radiation is in coherent
state, $\Gamma=1$ for all curves.}
\end{figure}

Oscillating behavior of the atom excitations (Rabi oscillations) can
be also seen in Fig. 4 which illustrate typical solutions of Eq.
(\ref{toh}). The most pronounced oscillations are for larger values
of $\Gamma$ and $N_0$.

The solution of Eq. (\ref{toh}) is also used to obtain a
configuration-space densities of transmitted and reflected photons.
The calculated data are shown in Fig. 5. Qualitatively, the curves
are similar to those in Fig. 2 if $N_0$ is small (see Fig. 5a).
Pronounced oscillations are present only for the reflected photons
when $N_0$ is sufficiently large (see the inset in Fig. 5b).

The numbers of reflected and transmitted photons are obtained after
integration of the photon densities, $\hat{\rho}_r$ and
$\hat{\rho}_l$,
\begin{equation}\label{tfh}
N_r=\langle \hat{N}_r\rangle =\int_{v(t_0-t)}^0dx\langle
\hat{\rho}_r(x,t)\rangle=\frac\Gamma 4\int_{t_0}^td\tau\langle
\Sigma_\tau \rangle,
\end{equation}
\begin{equation}\label{tfih}
N_l=\langle \hat{N}_l\rangle =\int_0^{v(t-t_0)}dx\langle
\hat{\rho}_l(x,t)\rangle= \int_{t_0}^td\tau v\langle
\tilde{\rho}_l[v(t-\tau),t]\rangle-N_r,
\end{equation}
where the conditions
$\tilde{r}_q|\Psi\rangle=\langle\Psi|\tilde{r}_q^ \dagger=0$ are
used. Intervals for integration over  $x$ are chosen to be
sufficiently large to cover the regions where the particle densities
differ from zero. Corresponding time interval, $t-t_0$, satisfies
condition (\ref{a}).

The term $-\partial_t \langle \sigma_z\rangle/2v$, which is
important for determining the pattern of the transmitted pulse, does
not contribute to the total number of the transmitted photons
because of zero value of $\langle \Sigma\rangle$ at the boundary
points $t$ and $t_0$. The calculated values of $N_r$ and $N_l$ are
shown in Fig. 6 by  dashed lines.

To estimate  upper limit of the reflected photon number, $N_r$, the
inequality  $\langle \Sigma\rangle<2$ and Eq. (\ref{tfh}) are used.
Thus we have $N_r\leq \Gamma w/(2v)$. The limiting value of $N_r$
does not depend on $N_0$ and can be small even if $N_0>>1$.
Therefore the reflected radiation can be used as a controllable
source of few-photon pulses. Also reflected photons can be useful,
for example, to check the atom state or obtain the interaction
parameter $g$.

\section{FLUCTUATIONS OF OUTGOING PHOTONS}

Since, according to Eqs. (\ref{thir}),(\ref{fift1}), and
(\ref{fift2}),  $\hat{N}_l+\hat{N}_r\equiv \tilde{N_l}$, noise
properties of the reflected and transmitted photons can be described
by the following variances
\begin{equation}\label{tsih}
\langle \Delta\hat{N}_r^2\rangle \equiv\langle
(\hat{N}_r-N_r)^2\rangle=\langle \hat{N}_r^2\rangle -N_r^2,
\end{equation}
\begin{equation}\label{tseh}
\langle \Delta\hat{N}_l^2\rangle \equiv\langle
(\hat{N}_l-N_l)^2\rangle=\langle
(\tilde{N}_l-\hat{N}_r)^2\rangle-N_l^2,
\end{equation}
where
\begin{equation}\label{tseh1}
\hat{N}_r=\int_{t_0}^t d\tau\bigg[\frac\Gamma4\Sigma +ig\int dq
(\sigma_+\tilde{r}_q-\tilde{r}_q^\dagger\sigma_-)\bigg]_\tau.
\end{equation}
 Eqs. (\ref{tsih}) and (\ref{tseh}) represent the mean square deviations of
the photon numbers from their average values $N_r$ and $N_l$. Since
there are no reflected photons when $g=0$, the fluctuations of the
transmitted photons are identical to those of the incident
radiation:
\begin{equation}\label{teih}
\langle \Delta\hat{N}_l^2\rangle=\langle
\Delta\tilde{N}_l^2\rangle=N_l=N_0.
\end{equation}
\begin{figure}[!ht]
\centering
\includegraphics[width=\textwidth]{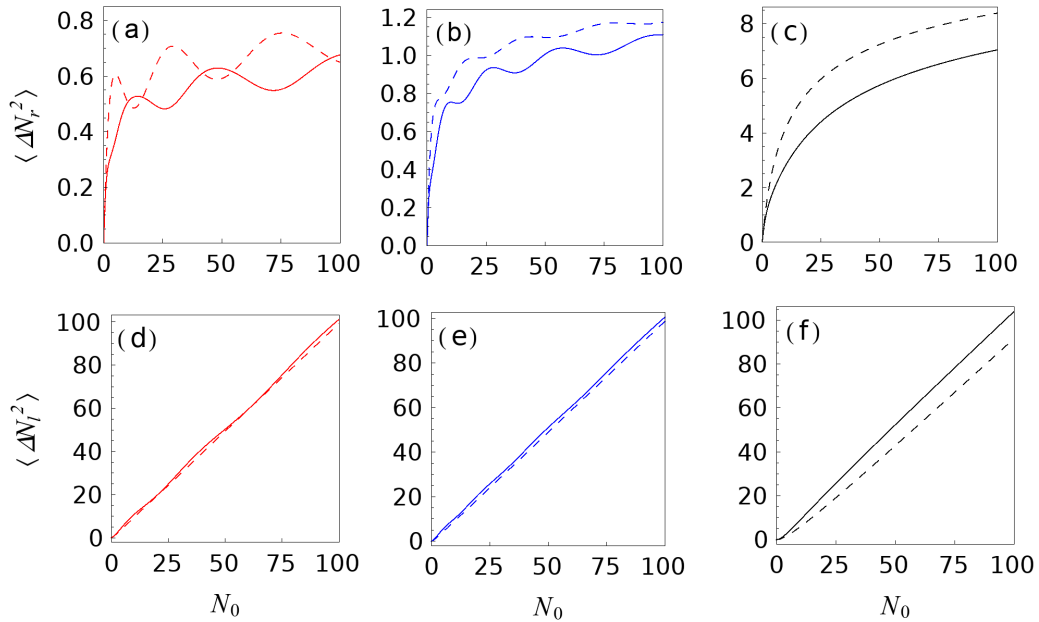}
\caption{(Color online) Photon number variances.  Solid lines -
obtained using the definitions (\ref{tsih}), (\ref{tseh}) and
numerical solutions of Eqs. (\ref{tenii}) and (\ref{tenj}). In the
case of Poissonian statistics of outgoing radiation, the
corresponding variances would be given by $N_{r,l}$ -  shown by
dashed lines. $\Gamma = 0.5$ for (a,d); $\Gamma = 1$ for (b,e);
$\Gamma =10$ for (c,f).}
\end{figure}
Eq. (\ref{teih}) can be easily verified  using the explicit term
(\ref{fiv}) for $\hat{N}_l$ and Eq. (\ref{tvsi}) for the multimode
coherent-state, $|\Psi \{\alpha\}\rangle$.

In what follows we study the modification of the photon statistics
caused by the radiation-atom interaction. To simplify further
analysis we again consider only the resonant case
$\omega_0=\omega_a$. As it is shown in the Appendix
\begin{equation}\label{tenh}
\langle\hat{N}_r^2\rangle=\frac{\Gamma^2}{8}\int_{t_0}^ t d\tau
\int_{t_0}^\tau d\tau^\prime\langle S(\tau,\tau^\prime)\rangle+N_r,
\end{equation}
 where
$S(\tau,\tau^\prime)=2\sigma_+(\tau^\prime)\Sigma(\tau)
\sigma_-(\tau^ \prime).$  It can be easily seen that the reflected
photons does not obey the Poissonian statistics if the first term in
the right side of Eq. (\ref{tenh}) is bigger or smaller than $N^2_r$
(super- or sub-Poissonian statistics, respectively).

The average value $\langle S(\tau,\tau^\prime)\rangle$ is governed
by the equation
\begin{equation}\label{tenii}
\hat{L}_\tau\langle S(\tau,\tau^\prime)\rangle=4g^2p^2(\tau)\langle
\Sigma(\tau^\prime)\rangle,
\end{equation}
which can be derived similarly to Eq. (\ref{toh}). The definition of
$\langle S(\tau,\tau^\prime)\rangle$ and properties of the Pauli
matrices, namely
$\sigma_+(\tau)\Sigma(\tau)=\Sigma(\tau)\sigma_-(\tau)=(\sigma_+)^2
=(\sigma_-)^2=0$, let us get the initial conditions for $\langle
S(\tau,\tau^\prime)\rangle$ as $\langle
S(\tau,\tau^\prime)\rangle=\partial_\tau\langle S(\tau,
\tau^\prime)\rangle=0$ when $\tau=\tau^\prime$.

After solving Eqs. (\ref{toh}) and (\ref{tenii}), it becomes
possible to calculate the mean square value $\langle
\hat{N}_r^2\rangle$ of the reflected photons. In Fig. 6a crossovers
from sub-Poissonian to super-Poissonian statistics are seen for some
specific values of $N_0$ . For bigger $\Gamma$ (see. Figs. 6b,c) the
variances $\langle \Delta\hat{N}_r^2\rangle$ are characterized by
sub-Poissonian statistics. Similar to results of Sect. 4 we see
nonclassical nature of outgoing radiation that can be used in
applications.

To obtain  $\langle \Delta\hat{N}_l^2\rangle$ we should know not
only $\langle \hat{N}_r^2\rangle$ but also
$\langle\tilde{N}_l\hat{N}_r\rangle=\frac\Gamma4\int_{t_0}^t d\tau
\langle \tilde{N}_l\Sigma_\tau \rangle $. The quantity $\langle
\tilde{N}_l\Sigma_\tau \rangle$ entering the integrand obeys the
equation
\begin{equation}\label{tenj}
\hat{L}_\tau\langle  \tilde{N}_l\Sigma(\tau)
\rangle=4g^2p^2(\tau)\langle N_0-\sigma_z(\tau)\rangle.
\end{equation}
Initial conditions are the same as for $\langle\Sigma(\tau)\rangle$,
i.e.
\[\langle \tilde{N}_l\Sigma(\tau=t_0)\rangle=\partial_\tau \langle
\tilde{N}_l\Sigma(\tau=t_0)\rangle=0.\]

Fluctuations of the transmitted radiation are very similar to the
fluctuations of the incident light (see Figs. 6d-f). This is due to
saturation of the TLS response: only insignificant number of photons
are involved in the atom excitation. Most photons are not affected
by the atom and conserve the statistical properties of the
coherent-state input.

\section{Discussion and Conclusion}
The purpose of this paper is to analyze  distinct features of
outgoing radiation. These features describe not only spatial but
also frequency distribution (i.e. spectrum) of the radiation.
Therefore it is appropriate to use the method of photon distribution
functions. Restricting our analysis to an incident pulse formed as
Gaussian packet of single-photon Fock state it becomes possible to
obtain analytical expressions for distribution functions of outgoing
photons. In Fig. 1 one can see the change of sign of distribution
function of the transmitted photons. This means that $\langle
f^{r,l}(x,q,t)\rangle$ describes rather a quasiprobability than a
probability of photon distribution.

Integrating $\langle f^{r,l}(x,q,t)\rangle$ over variables $q$ or
$x$ we obtain spatial or frequency distributions, respectively.
Spectra of transmitted and reflected radiation have very different
structures strongly dependent on both the detuning, $\omega_{a0}$,
and the dimensionless parameter $\Gamma w/v$ (see Fig. 3).

The criterion of negligible reflection as well as the criterion of
negligible transmission are derived using the explicit term for
spatial distribution of photons, Eq. (\ref{tvtw}). These criteria
and the data in Fig. 2 agree well with earlier studies in this field
which show a higher probability for short pulses to be transmitted.

The case of coherent state of the incident radiation is also
considered. The excitation-relaxation rates of the atom depend on
the number of photons, $N_0$. For multi-photon pulses, $N_0>>1$, the
Rabi frequency is proportional to $N_0^{1/2}$.  The tendency for
oscillation frequency to grow with $N_0$ is seen in Fig. 4.

Numerical data in Figs. 4, 5 are obtained from solution of Eq.
(\ref{toh}) which governs the evolution of $\langle
\Sigma(t)\rangle\equiv\langle\sigma_z(t)\rangle+1$ . The quantity
$\langle\Sigma(t)\rangle$ describes the probability of TLS to be
excited. At the same time, $\langle \Sigma(t)\rangle$ and its
derivative $\partial_t\langle \Sigma(t)\rangle$ determine the photon
densities $\langle \rho_{l,r}(x,t)\rangle$ [see Eqs.
(\ref{tvei},\ref{tvni})]. The interconnection of these physical
quantities is explained by  energy conservation: each atom
excitation is accompanied by annihilation of one photon in the
waveguide and vice versa [see the interaction term in Eq.
(\ref{one})].

In view of possible applications of outgoing radiation its noise
characteristics are also important. Eqs. (\ref{tseh}) and
(\ref{tenh}) and solutions of Eqs. (\ref{tenii},\ref{tenj}) make it
possible to calculate variances  of photon numbers. It can be seen
from Fig. 6 that in most cases the transmitted photons obey the
super-Poissonian statistics, while the reflected photons obey the
sub-Poissonian one that is in a qualitative consistence with results
of Ref. \cite{koca} where bunching of transmitted photons and
antibunching of reflected photons only were obtained. In our
formalism, the case considered in \cite{koca} corresponds to
infinitely long pulses. Hence, comparison of \cite{koca} with our
data can be plausible for only large values of the parameter $\Gamma w/v$
used in Fig. 6.

The approach used in the present paper can be easily modified to study more
complex phenomena.
Among of these, propagation of electromagnetic pulses in a waveguide,
coupled to a pair of TLS, is of interest.

\section{Acknowledgment}
We thank V. Bondarenko and A. Sokolov for their interest to this
research and stimulating discussions.

\section{Appendix: derivation of Eq. (\ref{tenh})}

It follows from Eq. (\ref{tseh1}) that the average $\langle
\hat{N}_r^2\rangle$ is given by
\begin{equation}\label{A1}
\langle \hat{N}_r^2\rangle =\frac {\Gamma^2}
{16}\int_{t_0}^td\tau\int_{t_0}^td\tau^\prime\Bigg [\langle \Sigma
(\tau)\Sigma (\tau^\prime )\rangle+\frac{iv}{\pi g}\int
dq\langle\sigma_+(\tau^\prime)\tilde{r}_q(\tau^\prime)\Sigma(\tau)-h.c.\rangle
\end{equation}
\[+\frac{4v}{\pi \Gamma}\int dq\int dq^\prime\langle
\sigma_+(\tau)\tilde{r}_q(\tau ){\tilde
r}^\dag_{q^{\prime}}(\tau^\prime)\sigma_-(\tau^\prime)
\rangle\Bigg].\] It is useful to represent the product
$\tilde{r}_q(\tau ){\tilde r}^\dag_{q^{\prime}}(\tau^\prime)$ in the
ordered form as $\tilde{r}_q(\tau ){\tilde
r}^\dag_{q^{\prime}}(\tau^\prime)={\tilde
r}^\dag_{q^{\prime}}(\tau^\prime)\tilde{r}_q(\tau
)+\delta(q-q^\prime)\exp[i\omega_q^r(\tau^\prime-\tau)]$. The part
of the last term in Eq. (\ref{A1}) that is proportional to
$\delta(q-q^\prime)$ gives a contribution to $\langle
\hat{N}_r^2\rangle$ equal to
\begin{equation}\label{A2}
\frac {\Gamma} {4}\int_{t_0}^td\tau\langle \Sigma(\tau)\rangle =N_r.
\end{equation}
The remaining part of the third term in Eq. (\ref{A1}),
\begin{equation}\label{A3}
\int dq\int dq^\prime\langle \sigma_+(\tau){\tilde
r}^\dag_{q^{\prime}}(\tau^\prime)\tilde{r}_q(\tau
)\sigma_-(\tau^\prime) \rangle ,
\end{equation}
is equal to zero. To prove this let us consider the term $\int
dq^\prime\langle\Psi| \sigma_+(\tau){\tilde
r}^\dag_{q^{\prime}}(\tau^\prime)$ where $|\Psi\rangle$ is given by
Eq. (\ref{tvsi}). Representing ${\tilde
r}^\dag_{q^{\prime}}(\tau^\prime)$ as
\begin{equation}\label{A4}
{\tilde
r}^\dag_{q^{\prime}}(\tau)e^{i\omega_{q\prime}^r(\tau^\prime-\tau
)}= \bigg [r^\dag_{q^{\prime}}(\tau)-ig\int_{t_0} ^\tau
d\tau^{\prime\prime}e^{i\omega_{q\prime}^r(\tau-\tau^{\prime\prime})}
\sigma_+(\tau^{\prime\prime})\bigg
]e^{i\omega_{q\prime}^r(\tau^\prime-\tau)}
\end{equation}
and taking into account that $r^\dag_{q^{\prime}}(\tau)$ commutes
with $\sigma_+(\tau)$, we have
\begin{equation}\label{A5}
\int dq^\prime\langle\Psi| \sigma_+(\tau){\tilde
r}^\dag_{q^{\prime}}(\tau^\prime)=\int
dq^\prime\langle\Psi|\bigg[r^\dag_{q^{\prime}}(\tau)\sigma_+(\tau)
\end{equation}
\[-ig\int_{t_0}
^\tau
d\tau^{\prime\prime}e^{i\omega_{q\prime}^r(\tau-\tau^{\prime\prime})}
\sigma_+(\tau)\sigma_+(\tau^{\prime\prime})\bigg
]e^{i\omega_{q^\prime}^r(\tau^\prime-\tau)}\] \[=ig\int
dq^\prime\int_{t_0}^\tau
d\tau^{\prime\prime}e^{i\omega_{q\prime}^r(\tau^\prime-
\tau^{\prime\prime})}\langle\Psi|[\sigma_+(\tau^{\prime\prime}),
\sigma_+(\tau)]\]\[=i\frac{2\pi
g}v\langle\Psi|[\sigma_+(\tau^\prime),
\sigma_+(\tau)]\theta(\tau-\tau^\prime).\] The last expression in
Eq. (\ref{A5}) is obtained after integration over $q^\prime$ and
$\tau^{\prime\prime}$. Also, we use here the condition
$\langle\Psi|\tilde{r}_{q^\prime}^\dag=0$.

The value of $ \int dq\tilde{r}_q(\tau
)\sigma_-(\tau^\prime)|\Psi\rangle$ can be obtained from Eq.
(\ref{A5}) by means of Hermitian conjugation and replacement $\tau
\longleftrightarrow \tau^\prime $:
\begin{equation}\label{A6}
\int dq\tilde{r}_q(\tau )\sigma_-(\tau^\prime)|\Psi
\rangle=-i\frac{2\pi g}v[\sigma_-(\tau^\prime),
\sigma_-(\tau)]|\Psi\rangle\theta(\tau^\prime-\tau).
\end{equation}
Inserting the last terms of Eqs. (\ref{A5}) and (\ref{A6}) into Eq.
(\ref{A3}) we get zero.

A similar procedure is used to simplify the second term in the
brackets of Eq. (\ref{A1}). Repeating the previous reasonings we
obtain the following relations (see also the Appendix in Ref.
\cite{dom}):
\begin{equation}\label{A7}
\int dq\tilde{r}_q(\tau^\prime )\Sigma(\tau)|\Psi
\rangle=i\frac{2\pi g}v[\sigma_-(\tau^\prime),
\Sigma(\tau)]|\Psi\rangle\theta(\tau-\tau^\prime),
\end{equation}
\begin{equation}\label{A8}
\int dq\langle\Psi|\Sigma(\tau)\tilde{r}^\dag_q(\tau^\prime
)=i\frac{2\pi g}v\langle\Psi|[\sigma_+(\tau^\prime),
\Sigma(\tau)]\theta(\tau-\tau^\prime).
\end{equation}
Using Eqs. (\ref{A7}),(\ref{A8}) we have
\begin{equation}\label{A9}
\frac {\Gamma^2}{16}\int_{t_0}^td\tau\int_{t_0}^t
d\tau^\prime\frac{iv}{\pi g}\int
dq\langle\sigma_+(\tau^\prime)\tilde{r}_q(\tau^\prime)\Sigma(\tau)-h.c.\rangle
\end{equation}
\[=-\frac{\Gamma^2}{16}\int_{t_0}^td\tau\int_{t_0}^\tau d\tau^\prime\langle
\Sigma(\tau)\Sigma(\tau^\prime)+\Sigma(\tau^\prime)\Sigma(\tau)-
4\sigma_+(\tau^\prime)\Sigma(\tau)\sigma_-(\tau^\prime)\rangle.\]
Finally, the overall contribution of three terms in brackets of Eq.
(\ref{A1}) results in Eq. (\ref{tenh}).

\newpage \parindent 0 cm \parskip=5mm
\end{document}